\documentclass[reprint, twocolumn, amssymb, amsmath, bibnotes, aps, prl, superscriptaddress,longbibliography]{revtex4-2} 

\usepackage{graphicx,comment}
\usepackage{float}
\usepackage{dcolumn}
\usepackage{bm}
\usepackage{braket}
\usepackage[dvipsnames]{xcolor}
\usepackage[colorlinks=true]{hyperref}
\definecolor{darkblue}{RGB}{33,33,137}
\definecolor{darkred}{RGB}{193,23,23}
\hypersetup{
    colorlinks=true,       
    linkcolor=blue,     
    citecolor=blue,    
    urlcolor=blue      
} 
\usepackage{natbib}
\usepackage[caption=false]{subfig}
\usepackage{color,soul}
\usepackage{tikz,fp}
\usepackage{tikz-cd}
\usepackage{environ}
\usepackage{upgreek} 
\usepackage{relsize}
\usepackage{wasysym}
\usepackage{graphicx}   
\usepackage{subcaption} 
\usepackage{ragged2e}
\usepackage[normalem]{ulem}
\tikzset{
    cross/.pic = {
    \draw[rotate = 45] (-#1,0) -- (#1,0);
    \draw[rotate = 45] (0,-#1) -- (0, #1);
    }
}

\begin{document}

\title{Random displacements in critical Rydberg atom arrays}
\author{Xingyu Li}
\thanks{These two authors contributed equally.}
\affiliation{
Institute for Advanced Study, Tsinghua University, Beijing 100084, China
}

\author{Shuyan Zhou}
\thanks{These two authors contributed equally.}
\affiliation{Department of Physics, Fudan University, Shanghai, 200438, China}
\affiliation{
Institute for Advanced Study, Tsinghua University, Beijing 100084, China
}

\author{Xue Chen}
\affiliation{
Institute for Advanced Study, Tsinghua University, Beijing 100084, China
}

\author{Chengshu Li}
\email{chengshu@mail.tsinghua.edu.cn}
\affiliation{
Institute for Advanced Study, Tsinghua University, Beijing 100084, China
}

\author{Hanteng Wang}
\email{hantengwang.physics@gmail.com}
\affiliation{
Institute for Advanced Study, Tsinghua University, Beijing 100084, China
}

\date{\today}

\begin{abstract}
Rydberg atom arrays promise high-fidelity quantum simulations of critical phenomena with flexible geometries. Yet experimental realizations inevitably suffer from disorder due to random displacements of atoms, leading to departures from the expected behavior. Here, we study how such positional disorder influences the Ising criticality. Since disorder breaks the $\mathbb{Z}_2$ symmetry, one might expect the system to flow to an infinite-strength disordered fixed point, erasing all nontrivial critical features in low spatial dimensions. Remarkably, we find instead that disorder in Rydberg systems is subjected to nontrivial local constraints, making the physics markedly different from systems with more conventional spatially short-range correlated or long-range correlated disorder. This leads to new classes of criticalities even at dimensions where conventional disorder would destroy criticality altogether. We then demonstrate as a consequence how a novel pseudo-criticality emerges in Rydberg atom chains of experimentally realistic scale, and show that the renormalization group flow is governed by a locally constrained $\mathbb{Z}_2$-breaking perturbation. Our findings uncover new disorder-driven phenomena and underscore the importance of carefully treating disorder effects in quantum simulators.
\end{abstract}

\maketitle
Rydberg atom arrays are powerful platforms for quantum simulation. They provide flexible spatial control, strong and tunable Rydberg-mediated interactions, and direct access to probing quantum many-body states~\cite{bernien2017probing,browaeys20,Ebadi2021,Browaeys2021,Semeghini2021,manovitz2024quantum}. This makes them particularly well-suited for studying quantum phase transitions and critical phenomena~\cite{CardyBook,SachdevBook}. Accessing quantum critical points is a key goal, where their scale-invariant correlations reveal universal physics governing quantum matter. Notable examples realized on Rydberg platforms include studies of quantum Ising transitions~\cite{keesling2019quantum,slagle2021microscopic,kalinowski2022bulk,fang2024probing,zhang2025observation}, confinement--deconfinement transitions~\cite{surace2020lattice,cheng2023variational,cheng2023gauge,cheng2024emergent}, and tricritical Ising points featuring emergent spacetime supersymmetry~\cite{Li2024,wang2025tricritical,naus2025measurement}.

However, disorder is an unavoidable experimental reality, arising naturally from inhomogeneities present in both condensed matter systems and quantum simulators~\cite{xing2015Griffiths,dag2024disorder,yue2025disorder}. Crucially, disorder is more than just a technical limitation. The interplay of disorder and interactions can drive novel phases of matter, including many-body localization~\cite{basko2006metal,Huse2013,Pollmann2014,Smith2016MBL}, spin glass phases~\cite{Parisi1983,mezard1987spin,Benjamin2024glass}, and percolation transitions~\cite{ShklovskiiBook,Skinner2019,Pant2019Percolation}. Critically, disorder also impacts quantum criticality itself. It can act as a relevant perturbation under renormalization group (RG) flow. This means disorder destabilizes clean critical points, altering the RG fixed points and thus changing the universal long-distance physics~\cite{harris1974effect,Halperin1983,altland2014anderson,Narovlansky2018,Lucas2023Disordered,liu2025deep,wang2025kink}. It can even induce novel disordered critical points with distinct universal behavior, such as those exhibiting Parisi--Sourlas supersymmetry in the random-field Ising model~\cite{parisi1979random,fisher1992random,fisher1995critical,Rychkov2022Parisi}. More broadly, understanding how disorder influences critical phenomena offers valuable insights into quantum optimization problems \cite{lucas2014ising,ebadi2022quantum,wang2022mbl,king2023quantum,Li2023optimization,Parisi2024nature,Qiu2025Annealing}.

\begin{figure}[t!]
    \centering
    \includegraphics[width=0.42\textwidth]{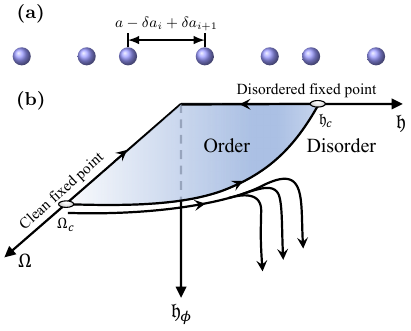} 
    \caption{\justifying (a) Schematic of a disordered Rydberg atom chain. Each atom is displaced from its designed position by $\delta a_i$, which is randomly distributed. (b) Schematic phase diagram and renormalization group flows of the disordered chain. $\Omega_c$ represents the clean criticality. A random $\partial \phi$ term drives the clean critical system to a novel disordered fixed point (on the $\mathfrak{h}$-axis), but the presence of higher-order random $\phi$ terms makes the system pseudo-critical. {That is, the system exhibits key signatures of a critical point over a wide intermediate scale, but is not a true critical point in the infinite-size limit.} The three different RG trajectories correspond to varying nearest-neighbor interaction strengths. With stronger interactions, the trajectory detours later.}
    \label{fig:main}
\end{figure}

In current Rydberg experiments, atoms are confined in optical tweezers, typically exhibiting positional deviations of about 5\% relative to the lattice constant \cite{fang2024probing}, {resulting mostly from imperfections to optical tweezers that trap the atoms. The dominant effect when the imperfections are small comes from random parallel-to-the-lattice displacements (along the lattice for $d=1$), as their first-order effect on interatomic distance outweighs the second-order effect of displacements perpendicular to the lattice.}  Motivated by this, we investigate the Ising transition realized in one-dimensional Rydberg atom arrays under such {longitudinal} quenched displacement disorder, see Fig.~\ref{fig:main}(a). In the absence of disorder, spontaneous breaking of a $\mathbb{Z}_2$ symmetry occurs and is effectively described by the celebrated $\phi^4$ theory, with $\phi$ as the order parameter [cf.~Eq.~\eqref{eq:phi4}]. When introducing disorder, the Rydberg blockade interactions lead to explicit breaking of the $\mathbb{Z}_2$ symmetry. Notably, we find that the effective action does not conform to the random-field Ising perturbation $S_\text{dis}=\int dxdt\,h(x)\phi(x,t)$ with a short-range correlated $h(x)$, which would destroy the ordered phase entirely \cite{imry1975random}. Instead, it is dominated by a locally constrained disorder term, $S_\text{dis} = \int dxdt\,h(x)\partial_x\phi(x,t)$, with $h(x)$ proportional to the random displacement at position $x$. This seemingly modest modification of the disorder action can drastically alter (i) the existence of the ordered phase and (ii) the universality class of the resulting disorder-driven criticality compared to the standard random-field Ising model, as detailed below.

\emph{Model.---}
We consider a one-dimensional chain of $N$ atoms. The atom at site $i$, with spatial position $x_i$, can be in the atomic ground state $|g_i\rangle$ or the Rydberg state $|r_i\rangle$. A laser couples these two states, determining the Rabi frequency $\Omega$ and detuning $\Delta$ through the laser strength and frequency, respectively. Two Rydberg atoms can interact via a power-law decaying interaction $V_{ij}=\frac{C_6}{|x_i - x_j|^6}$. The Hamiltonian reads
\begin{equation}
      H =  \frac{\Omega}{2} \sum_{i=1}^{N} {X}_i -\Delta \sum_{i=1}^{N} {n}_i + \sum_{i < j} V_{ij} {n}_i {n}_j
\end{equation}
where $ X_i=|g_i\rangle\langle r_i|+\text{h.c.}$ and $ n_i=|r_i\rangle\langle r_i|$. In the absence of disorder, the lattice constant is set to $a$.

We consider the regime of $C_6$ which achieves nearest-neighbor blockade, where the simultaneous excitation of two atoms in the low-energy sector is prevented. In our setup, we fix $C_6$ and $\Delta$ throughout to ensure a classical limit ($\Omega=0$) ground state in an ordered configuration, i.e., $|... grgrgr ... \rangle$ or $|... rgrgrg ... \rangle$. This breaks a $\mathbb{Z}_2$ symmetry, which is characterized by the lattice translation symmetry by one site. On the other hand, with a high Rabi frequency $\Omega$, this symmetry is restored in the ground state. At a certain critical value of the Rabi frequency $\Omega_c$, a 1+1D Ising critical point emerges. This is the ``starting point'' of our study that is located on the $\Omega$-axis in Fig.~\ref{fig:main}(b). 

Our main interest lies in how this phase transition is affected when small, time-independent random displacements in the atomic positions are introduced. To model this, we shift the position of each atom by $\delta a_i$ as shown in Fig.~\ref{fig:main}(a), where $\delta a_i$ are drawn independently from a uniform distribution on $[-\delta a, \delta a]$. The interaction term can then be Taylor expanded in small $\delta a_i$ as 
\begin{equation}
\label{eq:Taylor}
\begin{split}
         \sum_{i< j} V_{ij} {n}_i {n}_j&\approx\sum_{i=1}^N\sum_{l=1}^{l_0}\bigg[\frac{C_6}{(la)^6}{n}_i {n}_{i+l}\\
         &+\frac{6 C_6}{(la)^7}(\delta a_{i}-\delta a_{i+l}){n}_i {n}_{i+l}+\mathcal{O}(\delta a^2)\bigg],
\end{split}
\end{equation}
where we have assumed a periodic boundary condition $i\sim i+N$. We have also cut off the sum over $l$, which should have an upper limit of order $N$, to some constant upper limit $l_0$ independent of $N$.

\emph{Field Theory Treatment and Symmetries.---}
Critical phenomena exhibit universal long-distance physics, which should be insensitive to microscopic details. Here, we employ symmetry arguments to map our microscopic model onto an effective theory using conformal field theory (CFT)~\cite{DiFrancescoCFTBook}.

We begin by considering the clean system. The nearest-neighbor blockade model undergoes a $\mathbb{Z}_2$ breaking transition, described by the Ising universality class. This transition is captured by the $\phi^4$ field action
\begin{equation}\label{eq:phi4}
    S_0=\int d x dt \left[\frac12(\partial\phi)^2-\frac{m^2}2\phi^2-\lambda\phi^4\right],
\end{equation}
where $\phi$ is a real scalar field, the effective mass $m$ is tuned to criticality, and the $\mathbb Z_2$ symmetry is realized as $\phi\rightarrow-\phi$.

The scaling dimensions of the fields determine the critical exponents, which are experimentally measurable quantities of central interest. The two most relevant (i.e., lowest scaling dimensions) fields are $\phi$ and $\phi^2$, whose weights are both anomalous. The two-point correlation function of $\phi$ at criticality scales as $\langle \phi(x) \phi(0) \rangle \sim |x|^{-2 \Delta_\phi}$, where the scaling dimension $[\phi]  \equiv \Delta_\phi = 1/8$. Meanwhile, $\phi^2$, which couples to the effective mass $m$ and tunes the transition point, has the scaling dimension $[\phi^2]  = 1$.

With these preliminaries, we can now describe the effects of displacement disorder. A convenient way for treating disorder as a conformal field perturbation is as follows \cite{slagle2021microscopic,wang2025kink}: we first identify the symmetry of the microscopic disorder and then find the conformal fields that respect this symmetry. Among these fields, we select the lowest-weight one, as it dominates the long-distance physics near criticality. From the microscopic disorder Hamiltonian in Eq.~\eqref{eq:Taylor}, the first-order term yields $\sum_i \delta a_i {O}_i$, where ${O}_i = \sum_l ({n}_i {n}_{i-l} - {n}_i {n}_{i+l})$. The operator ${O}_i$ is odd under reflection about site $i$ ($l \mapsto -l$), or ${O}_i \mapsto -{O}_i$. The two lowest-weight primary fields $\phi$ and $\phi^2$, being even under reflection,  do not share this property. Hence, the celebrated random-field or random-bond Ising models do not fit the random Rydberg model we consider. On the other hand, the spatial derivative $\partial_x$ has odd parity under reflection and increases the scaling dimension by one. Therefore, the lowest-weight field compatible with the disorder symmetry is $\partial_x \phi$, with scaling dimension $9/8$. Numerical justification for this choice as the most relevant perturbation in the 1D critical Rydberg chain is provided in the Supplemental Material~\cite{supp}.

Thus, the disorder effective action can be expressed as
\begin{equation}\label{eq:partial}
\begin{aligned}
    S_\text{dis}[h]&=\int d x dt\;h(x)\partial_x\phi(x,t)\\
    &=\int d x dt\;(-\partial_xh(x))\phi(x,t),
\end{aligned}
\end{equation}
where the random quenched displacements exhibit short-range correlations 
\begin{equation}\label{eq:dis_corr}
    \overline{h(x) h(y)} = \mathfrak{h}^2 \delta(x - y),\,\text{and}\,\, \overline{h(x)}=0.
\end{equation}
Here, the overline denotes the disorder average, and $\mathfrak{h} \propto \delta a$ characterizes the typical deviation of atoms from their ideal positions. This type of random model indeed breaks the $\mathbb{Z}_2$ symmetry. When viewed as a random-field Ising model (as in the second line of Eq.~\eqref{eq:partial}), however, this action encodes that the perturbing longitudinal magnetic fields are constrained to locally sum to zero. Hence, we refer to this new type of disorder effect as the \emph{locally constrained} random model, which exhibits unexpected phases and criticalities, as detailed below.

\emph{Ordered Phase.---}
A crucial difference between a standard random field and a locally constrained random field lies in whether long-range order can survive.

For the random-field Ising model, Imry and Ma \cite{imry1975random} showed that order is unstable in one spatial dimension. Indeed, given a $d$-dimensional model $H_\text{Ising}+\sum_ih_iZ_i$, their argument proceeds as follows: Suppose a ferromagnetic ground state exists, and consider flipping all spins within a large region of linear size $L$ and volume $\mathcal{V}\sim L^{d}$. Two types of energy contributions arise and compete: (i) A domain wall of area $\sim L^{d-1}$ forms at the boundary, costing energy $E_{\mathrm{domain}}\sim J L^{d-1}$, where $J$ is the energy scale of $H_\text{Ising}$. (ii) Each flipped spin gains or loses energy of order $\mathfrak{h}$ from the local random field.  Summing over the $\mathcal{V}$ uncorrelated sites and applying the central limit theorem yields a contribution $E_{\mathrm{random}}\sim \mathfrak{h} \sqrt{L^{d}}$. For dimensions $d > 2$, we have $E_{\mathrm{domain}}\gg E_{\mathrm{random}}$ for large $L$, thus making the domain wall energetically costly and the ordered phase stable. In contrast, for $d=1$, the fluctuation dominates, destroying the assumed ordered phase.

On the other hand, the scenario changes for the locally constrained random field model, which can be modeled by $H_\text{Ising}+\sum_{\langle i,j\rangle} h_{ij}(Z_i-Z_j)$. Here, flipping all spins inside region $\mathcal{V}$ incurs no energy cost from the interior, as spins there remain nearly aligned, with the difference term $Z_i-Z_j\approx0$. Thus, the only energy gain arises at the boundary, with $E_{\mathrm{random}}\sim \mathfrak{h} \sqrt{L^{d-1}}$. Consequently, fluctuations never dominate and the ordered phase always persists for $d>1$. The same conclusion also holds for $d=1$ if the disorder field $h$ is bounded and small, since in this case the energy $E_{\mathrm{random}}$ is constrained to be bounded and small, and could not overcome the finite $E_{\mathrm{domain}}\sim J$~\footnote{{In $d=1$, the ordered phase would not exist if the random distribution is unbounded. In this case, extremely rare and unrealistically large defects could occur and introduce domain walls along the chain. We avoid this pathology by using a bounded distribution for the disorder.}}.

\begin{figure}[t]
    \centering
    \includegraphics[width=0.45\textwidth]{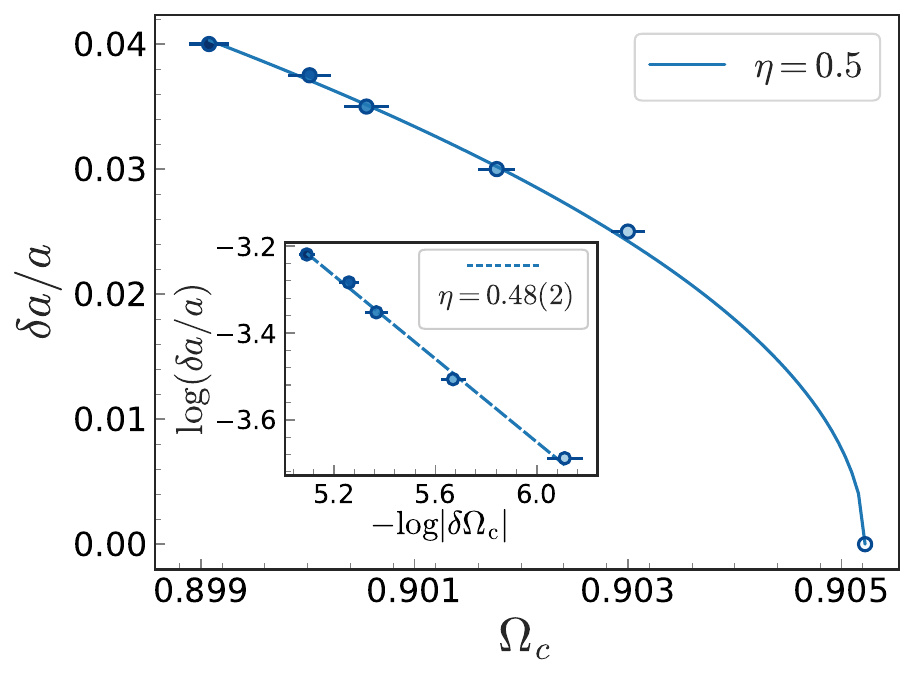} 
    \caption{\justifying Phase boundary of a disordered Rydberg chain for different displacement randomness $\delta a/a\propto\mathfrak{h}$. The interaction term is kept to the first order of $\delta a$ in numerics. Inset shows $\delta \Omega_c \propto \mathfrak{h}^{1/\eta}$ by a log--log scale plot, where $\delta \Omega_c(\mathfrak{h})=\Omega_c(\mathfrak{h})-\Omega_c(\mathfrak{h}=0)$ is the deviation of the critical Rabi frequency from the clean case due to finite disorder. {In numerical simulations, we choose $C_6=1.5^6$, $\Delta=1$, and $a=1$.}}
    \label{fig2}
\end{figure}

We numerically confirm an ordered phase in disordered Rydberg chains, which is captured by our locally constrained random-field model (Fig.~\ref{fig2}). We locate the phase boundary using the quotients method \cite{Fytas2013_L2L,Janus2013}, with the ground state obtained from density-matrix renormalization group (DMRG). Further computational details are given in the Supplemental Material~\cite{supp}.

\emph{Disordered Fixed Point.---}
The existence of an ordered phase suggests a possible phase diagram, as illustrated in Fig.~\ref{fig:main}(b). Importantly, this phase boundary should terminate at a point characterized by zero Rabi strength and finite disorder strength. We propose this as a novel disorder-controlled fixed point. Due to the absence of quantum fluctuations, this disordered critical point may be effectively described by a classical Hamiltonian
\begin{equation}\label{eq:dis_fix_point_model}
    H_\text{eff}=-J\sum_i Z_iZ_{i+1}+\sum_i h_i(Z_i-Z_{i+1}).
\end{equation}
For random strength $\tilde{\mathfrak{h}}$ smaller than the flipping energy cost $J$, the long-range ferromagnetic order remains. Here, the $\tilde{\mathfrak{h}}$ differs from $\mathfrak{h}$ in Eq.~\eqref{eq:dis_corr} by an order-one factor. However, once $\tilde{\mathfrak{h}}$ surpasses $J$, the local random fields $|h_i|$ can exceed $J$ at certain positions, flipping spins and fragmenting the ordered state. This defines a critical value $\tilde{\mathfrak{h}}_c=J$, separating ordered and disordered ground states. For $\tilde{\mathfrak{h}} > J$, the correlation length scales as $\xi \sim (\tilde{\mathfrak{h}} - J)^{-1}$. This is because the probability of finding a flipped spin nearby is proportional to $(\tilde{\mathfrak{h}}/J - 1)$. Translating this back to the disordered Rydberg chain yields $\xi \sim (\delta a - \delta a_c)^{-1}$, which is confirmed numerically.

Next, we address why we would expect the clean criticality to flow towards this disordered fixed point. We start from the Ising criticality described by Eq.~\eqref{eq:phi4} and introduce disorder via Eq.~\eqref{eq:partial} and Eq.~\eqref{eq:dis_corr} to examine its RG flow. Using perturbation theory, we calculate the disorder-averaged equal-time two-point correlator. Following Ref.~\cite{Aharony1976}, diagrammatically, the leading contributions are \cite{supp}
\begin{equation}
\overline{\langle \phi_k(t_0) \phi_{-k}(t_0)  \rangle}=
\begin{tikzpicture}[thick,scale = 0.3,baseline={([yshift=-2pt]current bounding box.center)}]
    \draw[thick] (0,0) -- (2.5,0);
\end{tikzpicture}
+
\begin{tikzpicture}[thick,scale = 0.3,baseline={([yshift=-2pt]current bounding box.center)}]
    \draw[thick] (0,0) -- (2,0);
    \draw[thick] (4,0) -- (6,0);
    \draw[densely dashed, thick] (2,0) -- (4,0);
    \path (2,0) pic[rotate = 0] {cross=4pt};
    \path (4,0) pic[rotate = 0] {cross=4pt};

\end{tikzpicture}
+ \cdots,
\end{equation}
where the non-disordered propagator with equal-time $t_0=0$ external legs in momentum space is

\begin{equation}\label{eq:clean_tree}
\begin{aligned}
\begin{tikzpicture}[thick,scale = 0.3,baseline={([yshift=-2pt]current bounding box.center)}]
    \draw[thick] (0,0) -- (2.5,0);
\end{tikzpicture}
&= G_k(t_0)
\propto \int d\omega\frac{1}{k^2-\omega^2}e^{i\omega t_0} \sim k^{-1}.
\end{aligned}
\end{equation}
For the disordered propagator, each cross represents a random field $h(x)$ coupling to $\partial_x \phi$, translated to momentum space as $k h_k \phi_k$. With dashed lines denoting disorder averaging $\overline{h_kh_{k'}}=\mathfrak{h}^2\delta(k+k')$, the result is
\begin{equation}\label{eq:dis_tree}
\begin{aligned}
\begin{tikzpicture}[thick,scale = 0.3,baseline={([yshift=-2pt]current bounding box.center)}]
    \draw[thick] (0,0) -- (2,0);
    \draw[thick] (4,0) -- (6,0);
    \draw[densely dashed, thick] (2,0) -- (4,0);
    \path (2,0) pic[rotate = 0] {cross=4pt};
    \path (4,0) pic[rotate = 0] {cross=4pt};
    \node at (2,1) {$h_k$};
    \node at (4,1) {$h_{k'}$};
    \node at (2,-1) {$t$};
    \node at (4,-1) {$t'$};
    \node at (0,-1) {$t_0$};
    \node at (6,-1) {$t_0$};
\end{tikzpicture}
&\propto \overline{\int dt[kh_kG_k(t)] \int dt'[k'h_{k'}G_{k'}(t')]} \\
&\propto \mathfrak{h}^2 \int d\omega\frac{k}{k^2-\omega^2}\frac{k}{k^2-\omega^2} \delta(\omega)\sim\mathfrak{h}^2k^{-2}.
\end{aligned}
\end{equation}
At long distances ($k \to 0$), the disorder-driven contribution Eq.~\eqref{eq:dis_tree} dominates over the clean propagator Eq.~\eqref{eq:clean_tree}. This dominance holds for higher-loop diagrams as well. Consequently, given diagrams allowed in clean action Eq.~\eqref{eq:phi4}, we could expect that the clean propagator \tikz[thick,scale = 0.3,baseline={([yshift=-2pt]current bounding box.center)}]{\draw[thick] (0,0) -- (2.5,0);}  can be replaced by a disordered propagator \tikz[thick,scale = 0.3,baseline={([yshift=-2pt]current bounding box.center)}]{\draw[thick] (0,0) -- (2,0);
    \draw[thick] (4,0) -- (6,0);
    \draw[densely dashed, thick] (2,0) -- (4,0);
    \path (2,0) pic[rotate = 0] {cross=4pt};
    \path (4,0) pic[rotate = 0] {cross=4pt};} 
to make diagrams more relevant at long distances, as long as the diagrams remain connected to external legs. 
Take the simplest loop diagram: It contains five clean propagators, of which at most three can be converted to disorder lines, for example
\begin{equation}
\label{eq:diagram}
\begin{aligned}
\begin{tikzpicture}[thick,scale = 0.3,baseline={([yshift=-2.5pt]current bounding box.center)}]
    \draw[thick] (-2.5,0) -- (6.5,0);
    \draw (2,0) circle(2);
\end{tikzpicture}
\longrightarrow
\begin{tikzpicture}[thick,scale = 0.3,baseline={([yshift=-3pt]current bounding box.center)}]
    \draw[thick] (-2.5,0) -- (0.6,0);
    \draw[densely dashed] (0.6,0) -- (3.3,0);
    \draw[thick] (3.3,0) -- (4.5,0);
    \draw[densely dashed] (4.5,0) -- (6.5,0);
    \draw[thick] (6.5,0) -- (7.5,0);
    \path (4.75,0) pic[rotate = 0] {cross=4pt};
    \path (6.5,0) pic[rotate = 0] {cross=4pt};
    \path (1,0) pic[rotate = 0] {cross=4pt};
    \path (3,0) pic[rotate = 0] {cross=4pt};
    \draw[-] (4, 0) arc (0:55:2);
    \draw[densely dashed] (3.16, 1.63) arc (55:135:2);
    \draw[-] (0.72, 1.52) arc (130:180:2);
    \draw[-] (0, 0) arc (180:360:2);
    \path (1,1.72) pic[rotate = 0] {cross=4pt};
    \path (3,1.72) pic[rotate = 0] {cross=4pt};
\end{tikzpicture}
\end{aligned}
\end{equation}
Hence we expect a new disorder-controlled fixed point is reached when all the diagrams are fully decorated by disorder propagators. Each dressed line comes with a $\delta(\omega)$  [cf. Eq.~\eqref{eq:dis_tree}], so all relevant diagrams are frozen at zero frequency. In other words, temporal direction drops out of the low-energy theory and the disordered criticality is purely classical. Based on this evidence, we conjecture that such a fixed point coincides with the transition point of the effective model in Eq.~\eqref{eq:dis_fix_point_model}.

\emph{Disorder-driven Scaling Laws.---}
Having established the existence of an ordered phase and disordered criticality, we now present several scaling laws that are experimentally relevant.

\begin{figure}[t]
    \centering
    \includegraphics[width=0.45\textwidth]{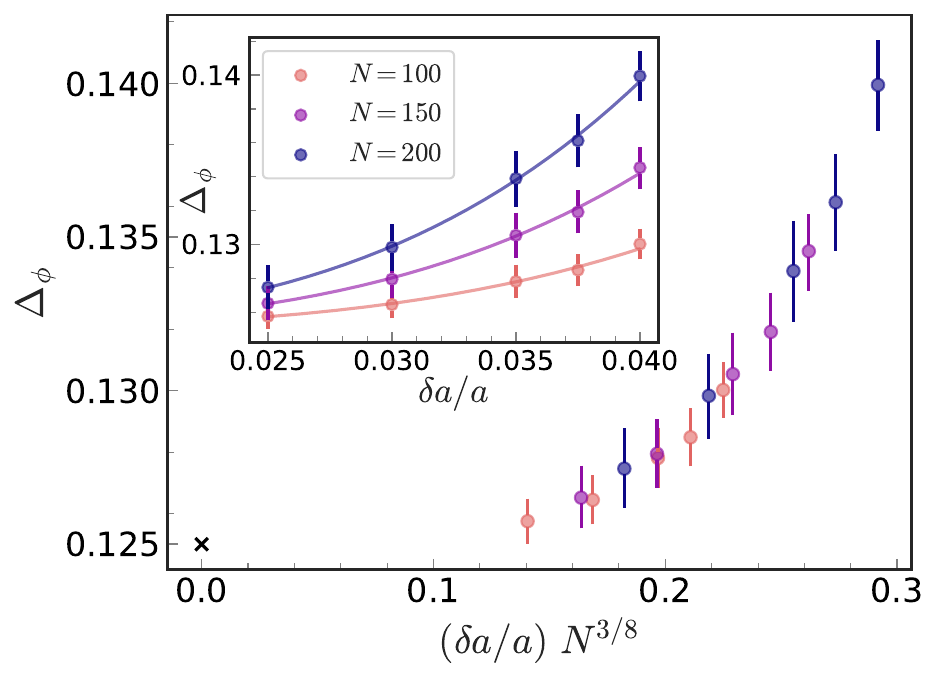}
    \caption{\justifying $\Delta_\phi$ with different displacement randomness $\delta a/a \propto \mathfrak{h}$ are extracted from two-point functions at finite size $N$. A finite-size scaling form of $\Delta_\phi(\mathfrak{h}, N) = f(\mathfrak{h} N^{3/8})$ is plotted in the main figure.}
    \label{fig3}
\end{figure}

In the clean critical case, the scaling dimension of the order parameter $\phi$ is known to be $\Delta_\phi = 1/8$. We investigate numerically how this scaling dimension evolves with increasing disorder strength $\mathfrak{h}$. The numerical results are shown in the inset of Fig.~\ref{fig3}. As $\mathfrak{h}$ grows, it deviates from the clean value of $1/8$, and this deviation occurs more rapidly with larger system sizes. This is consistent with expectations, as the RG flow towards disordered criticality at longer distances. Indeed, our numerical data follows a finite-size scaling law of the form $\Delta_\phi(\mathfrak{h}, N) = f(\mathfrak{h} N^{\gamma})$. To identify the exponent $\gamma$, we examine the dimension of $\mathfrak{h}$ dressed by disorder. We introduce replica indices $\alpha$ and $\beta$ to the random fields $h^\alpha(x)$ and $h^\beta(x)$, and then average over them. We obtain the effective action
\begin{equation}\label{eq:replica}
S_\text{dis}^{\alpha,\beta} = \mathfrak{h}^2 \int dx dt_1 dt_2\,\partial_x\phi^\alpha(x,t_1)\partial_x\phi^\beta(x,t_2).
\end{equation}
From this, one can immediately notice the dimension of the disorder strength to be $[\mathfrak{h}] = \gamma = \frac{1}{2}(3 - 2 - 2 \cdot \frac{1}{8}) = 3/8$, which is numerically confirmed in Fig.~\ref{fig3}.

We further explore another measurable quantity: the scaling behavior of the phase boundary as a function of disorder strength $\mathfrak{h}$. We define $\delta \Omega_c(\mathfrak{h})=\Omega_c(\mathfrak{h})-\Omega_c(\mathfrak{h}=0)$ as the deviation of the critical Rabi frequency from the clean case due to finite disorder. Since the Rabi frequency serves as a transverse field, its dimension is expected to match that of $[\delta \Omega_c] =[\phi^2] = 1$. Naively, one might anticipate $\delta \Omega_c \sim \mathfrak{h}^{8/3}$. However, numerical identification of the phase boundary reveals a different relation, closer to $\mathfrak{h}^2$, see Fig.~\ref{fig2}. To understand this discrepancy, a detailed analysis of Eq.~\eqref{eq:replica} is required. Divergences arise in Eq.~\eqref{eq:replica} when the replica indices coincide ($\alpha=\beta$) and the time arguments approach one another ($t_1 \rightarrow t_2$). These divergences require a counter-term and consequently shift the phase boundary. Using the fusion rule from CFT, we have~\footnote{This can be obtained from $\partial_x\phi(x,t_1)\partial_y\phi(y,t_2)\overset{x\rightarrow y,\,t_2\rightarrow t_1}{\sim}\text{const.}+\partial_x\partial_y ((x-y)^2+(t_1-t_2)^2)^{3/8}\phi^2(y)+\cdots$, where the terms we leave out are less singular.}
\begin{equation}
\partial_x\phi(x,t_1)\partial_x\phi(x,t_2)\overset{t_2\rightarrow t_1}{\sim}\text{const.}+\frac{\phi^2(x,t_1)}{|t_1-t_2|^{5/4}}+\cdots.
\end{equation}
Thus the most singular contribution in Eq.~\eqref{eq:replica} is
\begin{equation}
\label{eq:dissin}
S_\text{dis}^\text{singular} \sim \frac{\mathfrak{h}^2}{\mathfrak a^{1/4}} \int dxdt\,\phi^2(x,t),
\end{equation}
where a UV cutoff $\mathfrak a$ is introduced. This term acts as a mass term in the $\phi^4$ theory [cf.~Eq.~\eqref{eq:phi4}], shifting the phase boundary. Hence, this coefficient, proportional to $\mathfrak{h}^2$, acts as Rabi frequency in Rydberg case, leading to the observed relation $\delta \Omega_c \sim \mathfrak{h}^2$~\footnote{In regular random-field Ising models, this renormalization of the bare parameter $m^2$ does not occur, since in this case the perturbation is not heavy enough to induce a negative power of $\mathfrak a$ as in Eq.~\eqref{eq:dissin}. A direct power counting is therefore enough to locate the remnant of the phase boundary.}.

\emph{Pseudo-criticality.---}
In the discussion so far, we have only considered the first-order term of $\delta a$ in the Taylor expansion in Eq.~\eqref{eq:Taylor}. However, through direct calculation or symmetry considerations, we find that at second order, random $\phi$ perturbations appear in the action. Thus, the general structure of the disorder perturbation for the Rydberg chain becomes
\begin{equation}
    S_\text{dis} \sim \int dxdt \left[C_{1} \partial_x\phi \,(\delta a)^1+C_2 \phi \,(\delta a)^2+\cdots\right].
\end{equation}
Depending on the numerical values of $C_1$ and $C_2$, the trajectory of the RG flow from the clean to the disordered fixed point may or may not be interrupted by a flow out of the plane. For a conventional $S$-wave Rydberg interaction, which is isotropic and decays as $1/|x_i - x_j|^6$, we expect the flow to eventually leave the plane without ever reaching the disordered fixed point. Additionally, we anticipate that larger values of $C_1$ will cause the trajectory to move closer to the disordered fixed point, as shown in Fig.~\ref{fig:main}(b)~\footnote{For the conventional S-wave interaction in the nearest-neighbor blockade regime ($\frac{C_6}{a^6}\sim1$), we can estimate the fate of the RG flow as follows. When increasing the (dimensionless) length scale $l$ at which we make measurements, $C_1$ grows as $l^{3/8}$ by a power counting, while $C_2$ grows as $l^{11/8}$. By the scale $l_*$ where the $C_1$ term $\sim\frac{C_6}{a^6}\frac{\delta a}{a}{l_*}^{3/8}$ reaches order one, the $C_2$ term would have reached $\sim\frac{C_6}{a^6}\left(\frac{\delta a}{a}\right)^2{l_*}^{11/8}\sim\left(\frac{\delta a}{a}{l_*}^{3/8}\right)^2{l_*}^{5/8}\gg1$. This shows that while the $C_1$ term is more important at small scales, the $C_2$ term derails the RG flow before the disordered fixed point can be reached. This calculation also shows that the disruption occurs later if $\frac{C_6}{a^6}$ is larger.}.

Hence, in real experiments, the novel disordered criticality we propose is, in fact, pseudo-criticality. This presents an interesting scenario that can only be observed in scalable quantum simulators, while an inherently infinite-size system would not exhibit such behavior. We also note that for more exotic Rydberg interactions beyond the current $S$-wave type, there is a possibility that $C_2$ could be much smaller than $C_1$. We leave the exploration of such cases for future study.

{\emph{Tricritical Ising case.---}
In addition to the Ising universality class, we also examined disorder effects in the tricritical Ising (TCI), which is notable for its emergent spacetime supersymmetry in $1+1$ dimension \cite{Fendley2018,Li2024,cheng2025schwinger,wang2025tricritical,wang2025tricritical}. Such a TCI fixed point can emerge from a two-leg ladder of Rydberg atom arrays \cite{Li2024,wang2025tricritical}. Similar to the single-leg scenario, positional disorder induces a random perturbation of the form $\partial_x \phi$.}

{Performing numerical simulations of the two-leg Rydberg ladder model is computationally expensive when disorder averaging requires hundreds of realizations. Alternatively, we utilize a more computationally tractable model belonging to the same universality class, given by the  O'Brien--Fendley (OF) Hamiltonian~\cite{Fendley2018,slagle2021microscopic}:
\begin{equation}
\begin{split}
         H_\text{OF} =  &-\sum_{i} Z_{i} Z_{i+1} - g \sum_{i} X_{i}\\
         &+ \lambda \sum_{i}( X_{i-1} Z_{i} Z_{i+1} + Z_{i-1} Z_{i} X_{i+1}).
\end{split}
\end{equation}
Tricriticality emerges at $g = g_{c} = 1$ and $\lambda_{\text{TCI}} \approx 0.428$, where the scaling dimensions become $[\phi]=3/40$ and $[\phi^2]=1/5$.}

{We then examine how a locally constrained random field perturbation, represented by $\sum_i h_i(Z_i - Z_{i+1})$ with correlations $\langle h_i h_j \rangle = \mathfrak{h}^2\delta_{ij}$, impacts the TCI point. Notably, despite the changes in absolute scaling dimensions compared to the Ising case, we find that the phase boundary under this perturbation retains the same scaling behavior, i.e., $\delta g_c = g_{c}(\mathfrak{h}) - g_{c}(\mathfrak{h}=0) \propto \mathfrak{h}^2$, see Fig.~S2(c) in the Supplemental Material \cite{supp}. Because the TCI appears as a single point in the two-parameter phase diagram, quantifying its disorder-induced shift in transition position is crucial for experimental realization.}

\emph{Conclusion and Discussion.---}
In the one-dimensional disordered Rydberg chain, we uncover a new type of disorder effect, which is governed by random fields with local constraints. We propose its phase diagram and novel disordered criticality, supported by numerical verification. Additionally, we present preliminary results for higher dimensions (especially $d=2$) that suggest intriguing disorder-controlled fixed points, as detailed in Supplemental Material \cite{supp}. {In particular, we find suggestive evidence that upon the same uncontrolled diagram selection as before, an $\epsilon$-expansion from the upper critical dimension $d=4$ yields a Wilson-Fisher fixed point different from that of the Ising model.}

We briefly discuss how these disorder-driven scaling laws could be probed in experiments. Compared to preparing a critical ground state, ramping dynamics are straightforward to implement, yielding standard Kibble--Zurek scaling $\xi_\text{KZ} \sim N \mathcal{F}(s^\mu N)$ \cite{zhang2025observation,wang2025tricritical}. The introduction of a finite disorder strength $\mathfrak{h}$ adds an extra, relevant scale. Our theory predicts a generalized form $\overline{\xi_\text{KZ}} \sim N \mathcal{F}(s^\mu N, \mathfrak{h}^\gamma N)$, which could be directly tested in Rydberg platforms and would open a new avenue for exploring disorder-driven criticalities.

\emph{Note added.---}
Upon completing this project, we noticed a similar work on disordered Rydberg chains posted on arXiv \cite{sotogarcia2025randomness}. They discuss the Kibble--Zurek dynamics of the same setup as ours and identify a similar crossover regime due to finite size.

\emph{Acknowledgments.---}
We are grateful to Hui Zhai for stimulating ideas. This work is supported by National Natural Science Foundation of China under Grant No.~12504307 (C.L.). C.L. is also supported by Tsinghua University Dushi program. H.W. is supported by China Postdoctoral Science Foundation under Grant No.~2024M751609 and Postdoctoral Fellowship Program of CPSF under Grant No.~GZC20231364.
The DMRG calculations are performed using the ITensor library \cite{ITensor}.

\bibliography{Rydberg_disorder.bib}

\end{document}


\title{Supplemental Material for \\ ``Random displacements in critical Rydberg atom arrays''}
\author{Xingyu Li}
\thanks{These two authors contributed equally.}
\affiliation{
Institute for Advanced Study, Tsinghua University, Beijing 100084, China
}

\author{Shuyan Zhou}
\thanks{These two authors contributed equally.}
\affiliation{Department of Physics, Fudan University, Shanghai, 200438, China}
\affiliation{
Institute for Advanced Study, Tsinghua University, Beijing 100084, China
}

\author{Xue Chen}
\affiliation{
Institute for Advanced Study, Tsinghua University, Beijing 100084, China
}

\author{Chengshu Li}
\email{chengshu@mail.tsinghua.edu.cn}
\affiliation{
Institute for Advanced Study, Tsinghua University, Beijing 100084, China
}

\author{Hanteng Wang}
\email{hantengwang.physics@gmail.com}
\affiliation{
Institute for Advanced Study, Tsinghua University, Beijing 100084, China
}

\maketitle

\section{Justification of locally constrained model}
To test whether our locally-constrained effective theory captures the physics of the disordered Rydberg chain on mesoscopic scales, we examine an observable that cleanly separates a random gradient perturbation, $\partial_x\phi$, from a random field, $\phi$ perturbation. Specifically, we consider the connected two-point function 
\begin{equation}
    G(x_i-x_j) = \langle Z_i Z_j\rangle-\langle Z_i\rangle \langle Z_j\rangle.
\end{equation}

In the clean, finite-size chain, this correlator decays algebraically with scaling dimension $\Delta_\phi$ as $G(x) \propto \left[\sin(\pi x/N)\right]^{-2\Delta_\phi}$, by which $\log G(x)$ is linearly related to $\log\sin(\pi x/N)$ with slope $-2\Delta_\phi = -1/4$. Disorder breaks this global form, so we introduce a \emph{local} scaling dimension for the disorder-averaged correlator via the point-wise slope
\begin{equation}
    \Delta_\phi(\mathfrak{h},x) \equiv \frac{\partial \log \overline{G(\mathfrak{h},x)}}{\partial\log \sin(\pi x/N)},
\end{equation}
where the overline denotes disorder averaging.

\begin{figure}[t]
    \centering
    \includegraphics[width=0.7\textwidth]{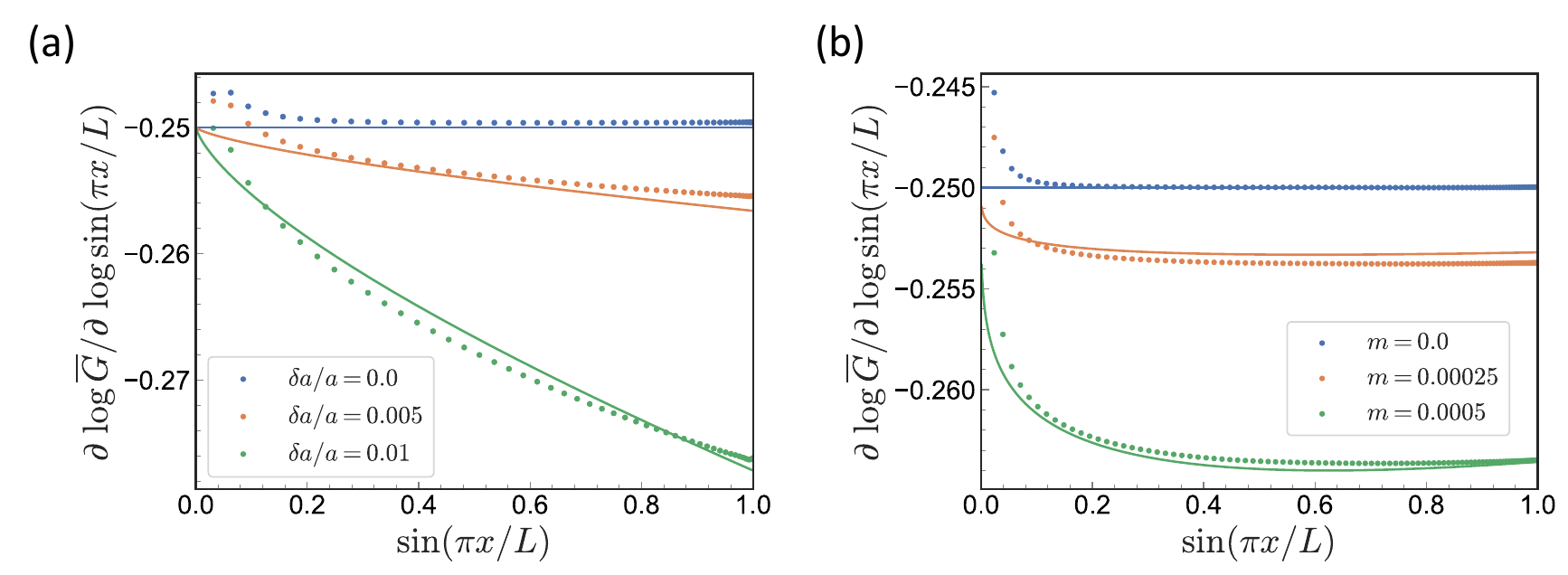} 
    \caption{\justifying (a) Dots: DMRG calculation of scaling behavior of disorder-averaged two-point function $\overline{G(x)}$ for the Rydberg chain with positional disorder retained to all orders in~$\delta a$. Solid line: Ising CFT result with locally-constrained disorder, using a fitted disorder strength. (b) Dots: DMRG calculation of scaling behavior of disorder-averaged two-point function $\overline{G(x)}$ with random $\phi$ disorder. Solid line: Ising CFT result with random $\phi$ disorder, using a fitted disorder strength.}
    \label{fig:compare}
\end{figure}

Figure~\ref{fig:compare}(a) shows DMRG data (dots) for the Rydberg chain with positional disorder retained to \emph{all} orders in~$\delta a$, together with field-theory calculations (solid curves) for an Ising chain subject to a random $\partial_x\phi$ term. The excellent agreement at small $\delta a$ confirms that gradient disorder indeed dominates on the finite length scales we probe.

In contrast, Fig.~\ref{fig:compare}(b) displays results for a random-$\phi$ perturbation with variance $\mathfrak{m}^2$, which compares DMRG numerics (dots) with field-theory calculations (solid curves). It is clear that the disordered Rydberg chain in Fig.~\ref{fig:compare}(a) bears no resemblance to the random-field Ising model in Fig.~\ref{fig:compare}(b), even though in the thermodynamic limit the true random field will ultimately take over\footnote{{This can be most clearly seen from the fact that both the lattice numerics and the field theory result in Fig.~\ref{fig:compare}(a) are lines with a significantly nonzero slope when $\sin(\pi x/L)$ is comparable to $1$. In fact, the slope is comparable to the slope of the line that joins the rightmost point on the line to the leftmost. In Fig.~\ref{fig:compare}(b), in contrast, all results are flat when $\sin(\pi x/L)$ is comparable to $1$. Thus a fit using the wrong model results in hideous disagreements.}}. For small disorder and experimentally relevant system sizes, the first-order $\delta a$ perturbation is the dominant effect, providing a solid foundation for our first-order approximation in the main text.

In both of the above figures, the solid line for theoretical comparison is obtained through a second order perturbative calculation on the Ising CFT, with one fitting parameter $\mathfrak{h}^2$ for the perturbation strength. The perturbed connected two-point correlator is 
\begin{equation}
\begin{aligned}
    \overline {G(x)}=\overline{\langle \phi(0)\phi(x)\rangle-\langle \phi(0)\rangle\langle\phi(x)\rangle}&=\int\mathcal Dh\Bigg[\frac{\langle \sigma(0,0)\sigma(x,0)e^{-\int dyd\tau\,h(y) O(y,\tau)}\rangle_\text{Ising}}{\langle e^{-\int dyd\tau\,h(y) O(y,\tau)}\rangle_\text{Ising}}\\&-\frac{\langle \sigma(0,0)e^{-\int dyd\tau\,h(y) O(y,\tau)}\rangle_\text{Ising}\langle\sigma(x,0)e^{-\int dyd\tau\,h(y) O(y,\tau)}\rangle_\text{Ising}}{\langle e^{-\int dyd\tau\,h(y) O(y,\tau)}\rangle_\text{Ising}^2}\Bigg],
\end{aligned}
\end{equation}
where on the left-hand side we compute expectations on the disordered ground state on a periodic line of length $L$ and then disorder average, and all operators are taken at the same time. The $\langle ...\rangle_\text{Ising}$ on the right-hand side denotes the expectation value on the Ising CFT on a cylinder of length $L$ and infinite imaginary time, and we have renamed $\phi$ to the more standard $\sigma$. We use the perturbation $O(y,\tau)=\partial_y\sigma(y,\tau)$ for Fig.~\ref{fig:compare}(a) and $O(y,\tau)=\sigma(y,\tau)$ for Fig.~\ref{fig:compare}(b). To lowest order in $\mathfrak{h}^2$ this reduces to
\begin{equation}
\begin{aligned}
    &\overline {G(x)}\approx\langle \sigma(0,0)\sigma(x,0)\rangle_\text{Ising}+\frac{\mathfrak{h}^2}{2}\int dyd\tau d\tau'\big[\,\langle \sigma(0,0)\sigma(x,0)O(y,\tau)O(y,\tau')\rangle_\text{Ising}\\&-\langle \sigma(0,0)\sigma(x,0)\rangle_\text{Ising}\langle O(y,\tau)O(y,\tau')\rangle_\text{Ising}-2\langle \sigma(0,0)O(y,\tau)\rangle_\text{Ising}\langle \sigma(x,0)O(y,\tau')\rangle_\text{Ising}\big].
\end{aligned}
\end{equation}
The integral involving the standard two- and four-point functions is done numerically to obtain the figures, where for completeness we include, for $L=2\pi$,
\begin{equation}
    \langle\sigma(x,\tau)\sigma(y,\tau')\rangle_\text{Ising}=\frac{e^{(\tau+\tau')/8}}{|e^{i x+\tau}-e^{i y+\tau'}|^{\frac14}}
\end{equation}
\begin{equation}
    \langle\sigma(x_1,\tau_1)\sigma(x_2,\tau_2)\sigma(x_3,\tau_3)\sigma(x_4,\tau_4)\rangle_\text{Ising}=\frac{e^{\sum_{i=1}^4 \tau_i/8}}2\left|\frac{z_{13}z_{24}}{z_{12}z_{23}z_{34}z_{41}}\right|^{\frac14}\big(|1+\sqrt{1-x}|+|1-\sqrt{1-x}|\big)
\end{equation}
In the last equation, $z_i=e^{i x_i+\tau_i}$, $z_{ij}=z_i-z_j$, and $x=\frac{z_{12}z_{34}}{z_{13}z_{24}}$ is the conformal cross ratio for the four-point function.

\section{Determination of the transition points}
In this section we outline our procedure for extracting the critical Rabi frequency. We first use the Binder-ratio crossing method. Defining the Binder ratio as
\begin{equation}
  U = \frac{\overline{\langle M^4 \rangle}}{\overline{\langle M^2 \rangle}^2}, \text{ and } M=\sum_i (-1)^i Z_i.
\end{equation}
By tuning the Rabi frequency \(\Omega\) in the Rydberg chain, the curves $U(\Omega)$ for different system sizes $N$ should ideally intersect at a single point, giving $\Omega_{U\text{-cross}}=\Omega_c$. To account for the $1/N$ effect, we employ the quotients method~\cite{Fytas2013_L2L,Janus2013}: for each pair of sizes $(N,bN)$ with fixed $b>1$, we locate the finite-size crossing $\Omega_{U\text{-cross}}(N)$ and assume $\Omega_{U\text{-cross}}=\Omega_c+A_UN^{-\alpha}$ where $\alpha$ is an unknown exponent.  In the presence of disorder, fitting this form alone leaves both $\Omega_c$ and $\alpha$ poorly constrained.

\begin{figure}[t]
    \centering
    \includegraphics[width=1\textwidth]{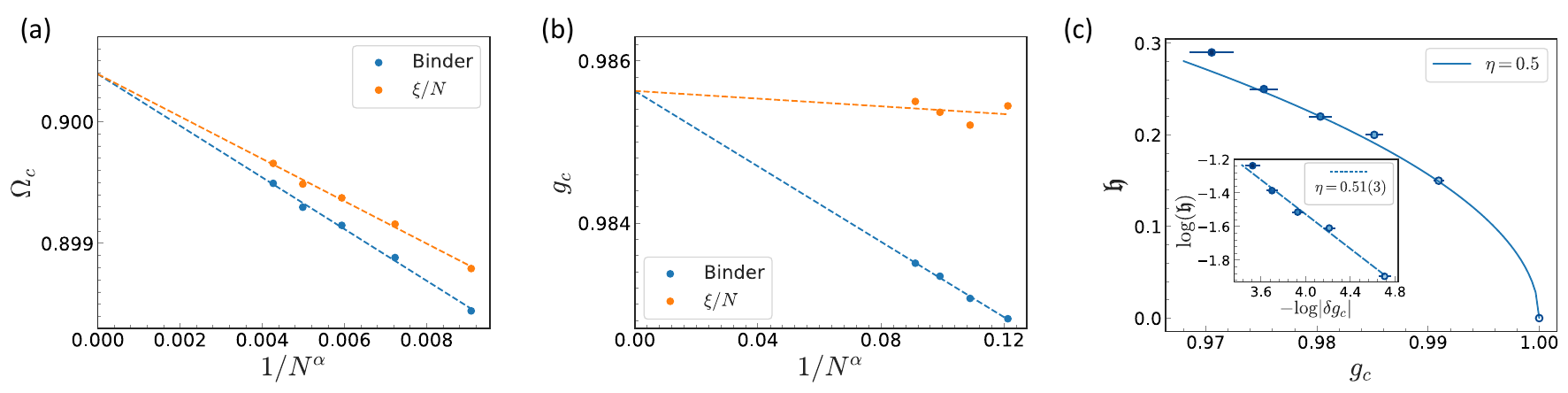} 
    \caption{\justifying (a) Computation of ${\Omega}_{c}$ for Rydberg system using the quotients method. We fit the crossing points ${\Omega}^{\text{cross }} (N)$ of lattice pairs $(N,3N/2)$ to the scaling form ${\Omega}^{\text{cross }} ={\Omega}_{c} +AN^{-\alpha}$, using both $U_{4}$ and $R_{\xi } =\xi /N$ to determine ${\Omega}^{\text{cross }}$. The data includes system sizes $(24,36),(28,42),(32,48),(36,54)$, and $(40,60)$, with disorder strength $\delta a/a =0.025$. (b) Computation of $g_c$ for TCI using the quotients method, similar to Ising case. The data includes system sizes $(14,21),(16,24),(18,27)$, and $(20,30)$, with disorder strength $\mathfrak{h}=0.2$. (c) Phase diagram near the critical point of the TCI, similar to Fig.~2. In the main figure, we use the theoretical prediction of exponent $0.5$. The inset shows a linear fit of the log--log plot, yielding a fitted exponent of $0.51 \pm 0.03$. }
    \label{fig:SM2}
\end{figure}

To improve the precision, we introduce a second dimensionless quantity, the normalized correlation length $R=\xi/N$. For a chain of length $N$, we compute
\begin{equation}
  \xi = \frac{1}{2 \sin\bigl(2\pi / N\bigr)} \,
  \sqrt{\frac{\chi(0)}{\chi\bigl(4\pi / N\bigr)} - 1}, \text{ where }\chi(k) =\sum_{\text{even }j} 
  G(x_j)\, e^{i k x_j}.
\end{equation}
Similarly, the crossings of $R(\Omega)$ occur at $\Omega_{R\text{-cross}}=\Omega_c+A_RN^{-\alpha}$ with the same exponent $\alpha$.  We therefore perform a \emph{joint fit} of both $\Omega_{U\text{‐cross}}(N)$ and $\Omega_{R\text{-cross}}(N)$ under the common parameters $\{\Omega_c,\alpha\}$.  Results for a fixed disorder strength are shown in Fig.~\ref{fig:SM2}(a).

\section{Results for the tricritical Ising case}
Beyond the conventional Ising criticality, Rydberg arrays can also exhibit tricritical Ising (TCI) points characterized by emergent spacetime supersymmetry, as proposed in Refs.~\cite{Li2024,wang2025tricritical}. Similarly to the standard Ising scenario, positional disorder induces a random perturbation of the form $\partial_x \phi$. In this section, we explore how such randomness influences the TCI point.

However, performing numerical simulations of the two-leg Rydberg ladder model introduced in Refs.~\cite{Li2024,wang2025tricritical} is computationally expensive when disorder averaging requires hundreds of realizations. To address this challenge, we utilize a more computationally tractable model belonging to the same universality class, given by the O'Brien--Fendley (OF) Hamiltonian~\cite{Fendley2018,slagle2021microscopic}:
\begin{equation}
H_\text{OF}=  -\sum_{i} Z_{i} Z_{i+1} - g \sum_{i} X_{i} + \lambda \sum_{i}( X_{i-1} Z_{i} Z_{i+1} + Z_{i-1} Z_{i} X_{i+1}).
\end{equation}
Here, the first two terms define the standard transverse-field Ising model, which exhibits Ising criticality at $g = g_{c} = 1$ and $\lambda = 0$. As $\lambda$ is increased to the tricritical value $\lambda_{\text{TCI}} \approx 0.428$, the model exhibits tricriticality. Within the range $0 < \lambda < \lambda_{\text{TCI}}$, standard Ising criticality persists, characterized by scaling dimensions identical to those discussed in the main text, i.e. $[\phi]=1/8$ and $[\phi^2]=1$. At the tricritical point ($\lambda = \lambda_{\text{TCI}}$), the scaling dimensions shift to $[\phi]=3/40$ and $[\phi^2]=1/5$.

We focus particularly on how a locally constrained random field perturbation, represented by $\sum_i h_i(Z_i - Z_{i+1})$ with correlations $\langle h_i h_j \rangle = \mathfrak{h}^2\delta_{ij}$, impacts the tricritical Ising point. Specifically, we investigate how the clean critical transition at $g_{c}$ shifts to $g_{c}(\mathfrak{h})$ in the presence of disorder. The transition positions are also calculated by the quotients method, see Fig.~\ref{fig:SM2}(b).  Notably, despite the changes in absolute scaling dimensions, we find that the phase boundary under this perturbation retains the same scaling behavior as derived in the main text, i.e. $\delta g_c = g_{c}(\mathfrak{h}) - g_{c}(\mathfrak{h}=0) \propto \mathfrak{h}^2$. This result is numerically verified and illustrated in Fig.~\ref{fig:SM2}(c).

\section{Higher dimensions}
To better understand how Rydberg systems deviate from their clean critical points once disorder is introduced, we will study this kind of disorder in more generality. Although we only considered the disordered Rydberg chain in the main text, the Taylor expansion and symmetry analysis also applies to higher dimensional settings. In a $d$ dimensional square lattice, where each atom is randomly displaced by a distance $\delta\vec a$ in these $d$ dimensions, to first order in $\delta\vec a$ and in each primary family, the lightest operators that couple to the displacements will be of the form $\vec\partial  O$ where $ O$ is some local operator. For concreteness and simplicity, in this section we will focus on special points in these models, where a clean system would be at a $d+1$-dimensional Ising critical point. More explicitly, we consider the model
\begin{equation}
    S[\vec h]=\int d^d\vec x dt\;\frac12(\partial\phi)^2-\frac{m_0^2}2\phi^2-\lambda_0\phi^4+\vec h\cdot\vec\partial\phi
\end{equation}
where $\phi$ is a real scalar field, $m_0$ and $\lambda_0$ are parameters tuned such that the system is at a $d+1$ dimensional Ising critical point when $\vec h=0$, and $\vec h$ is a $d$ dimensional random vector field whose distribution satisfies 
\begin{equation}
    \overline{h_i(t,\vec x) h_j(t',\vec y)}=\mathfrak h^2\delta_{ij}\delta(\vec x-\vec y)
\end{equation}
where $\mathfrak h^2$ is the control parameter for the disorder strength, and $i,j$ takes values in the spatial dimensions $1,...,d$.

We first justify our interest in the time-independent disorder by noting that at these critical points, the displacements that are randomly varying in time as well as in space, $\overline{h_i(t,\vec x) h_j(t',\vec y)}=\mathfrak h^2\delta_{ij}\delta(t-t')\delta(\vec x-\vec y)$, are not very interesting. This follows from the Harris criterion, which states that these disorders are irrelevant if (and only if) the weight of the operator coupling to disorder $x=[\vec\partial\phi]$ satisfies $x>\frac{d+1}2$. Since this is true at any nontrivial fixed point where $\phi$ has a positive anomalous dimension, or whenever $\phi$ is not free, these time-dependent random fluctuations for the atom positions are not very harmful for the measurement of the Ising critical properties.

On the other hand, the time-independent disorder we considered above is Harris relevant for $d<3$. To better understand the effect of this perturbation, we discuss an epsilon expansion near the upper critical dimension following the same type of diagram selection as in the main text.

A very simple situation occurs when $\lambda=0$, where the theory is free. We then obtain the exact result
\begin{equation}
    \overline{\langle\phi(\vec k,\omega)\phi(\vec k',\omega')\rangle_c}=\delta(\vec k-\vec k')\delta(\omega-\omega')\frac1{k^2-\omega^2}
\end{equation}
\begin{equation}
    \overline{\langle\phi(\vec k,\omega)\rangle\langle\phi(\vec k',\omega')\rangle}=\delta(\vec k-\vec k')\delta(\omega-\omega')\frac{\mathfrak h^2\delta(\omega)}{k^2}
\end{equation}
Restricting to slowly-varying observables in space and time, we see that similarly to the more familiar random-field Ising model, the disconnected part of the two-point correlator dominates over the connected part. In the following, we will therefore focus on the disconnected part of the correlators.

We next add a small $\lambda_0$ to the action and treat the problem perturbatively. As explained in the main text, ``internal lines" containing a disorder average have one less power of $k,\omega$ than the ones that do not, and when restricted to slowly-varying observables, this probably makes diagrams containing the most possible disorder averages at each loop order the most important~\cite{Aharony1976}. Retaining only this kind of diagram where $\omega=0$ in all lines, the physics of slowly-varying observables are controlled by that of the zero-frequency fields, which is consistent with our conjectured phase diagram. 

Indeed, focusing on these diagrams we find that the upper critical dimension of the problem is $d=4$ just as in a clean Ising theory in $d$ spacetime dimensions. The critical properties, at least of the two-point function, can then be computed through an $\epsilon$-expansion as in the standard $\phi^4$ theory. Nevertheless, a crucial distinction exists that makes physics in derivative-disordered systems different. This arises from the tension between the need to maximize the number of disordered internal lines and the need to keep the diagram connected. For example, the simplest two-loop self-energy correction to a propagator (as shown in Eq.~(13)) receives a combinatorial factor of $3$ from the three possible ways of keeping the self-energy connected, while the simplest one-loop correction to the $2-2$ interaction vertex receives a combinatorial factor of $2$ for the two choices of making one of the internal lines disordered. Therefore, the $\epsilon$-expansion and the associated Wilson--Fisher fixed point for derivative-disordered systems differ from the clean case by these combinatorial factors, and represent some unconventional critical phenomena.

\bibliography{Rydberg_disorder.bib}